\documentclass[preprint,12pt]{aastex}

\begin{document}
\shorttitle{Semimajor Axis Distribution of Extrasolar Planets}
\shortauthors{Currie, T. et al.}
\title{On the Semimajor Axis Distribution of Extrasolar Gas Giant Planets: Why Hot Jupiters Are Rare Around 
High-Mass Stars}
\author{Thayne Currie\altaffilmark{1}}
\altaffiltext{1}{Harvard-Smithsonian Center for Astrophysics, 60 Garden St. Cambridge, MA 02140}
\email{tcurrie@cfa.harvard.edu}
\begin{abstract}
Based on a suite of Monte Carlo simulations, I show that 
a stellar-mass-dependent lifetime of the gas disks from which planets form 
can explain the lack of hot Jupiters/close-in giant planets around high-mass 
stars and other key features of the observed semimajor axis distribution of 
radial velocity-detected giant planets.  Using reasonable parameters for the Type II migration rate, 
regions of planet formation, and timescales for 
gas giant core formation, I construct synthetic distributions of Jovian planets.
A planet formation/migration model assuming a stellar mass-dependent gas disk lifetime 
 reproduces key features in the observed distribution by preferentially 
stranding planets around high-mass stars at large semimajor axes.
\end{abstract}
\keywords{Stars: planetary systems: formation}
\section{Introduction}
The semimajor axis distribution of Jovian-mass planets discovered in 
radial velocity surveys reveals striking trends (Figure \ref{fg1}, top panels).
Many Jovian planets around solar-mass stars have 
semimajor axes a$_{p}$ $\gtrsim$ 0.5 AU.   Many also have a$_{p}$ $\lesssim$ 0.1--0.2 AU 
('hot Jupiters'), and few have intermediate values ('the period valley', \citealt{Cumming08}).
  While hot Jupiters comprise $\approx$ 20\% of 
planets around $<$ 1.5 M$_{\odot}$ stars, surveys have yet to
detect hot Jupiters orbiting $>$ 1.5 M$_{\odot}$ stars.  All radial velocity-detected 
planets around $>$ 1.5 M$_{\odot}$ stars have semimajor axes $\gtrsim$ 0.5 AU \citep{Jj07a, Jj07b, Jj08, Sa08, Wright09}.

Planetary migration may explain aspects of the semimajor axis 
distribution of Jovian planets, including the origin of hot Jupiters (which cannot 
form in situ ) and the period valley for solar-mass stars \citep[e.g.][]{Il04,Bi07}.
However, the cause for the lack of hot Jupiters and other close-in giant planets 
around high-mass stars is less clear.
One possibility is that hot Jupiters surrounding high-mass stars 
are engulfed as the stars evolved off the main sequence \citep{Sa08}.  However, most 
 high-mass stars with planets (subgiants) are physically too small to engulf hot 
Jupiters \citep{Jj07b}.  

In this paper, I show that models of planet formation/migration with a stellar-mass-dependent lifetime of the gaseous
circumstellar disks from which planets form can explain the dearth of 
hot Jupiters around high-mass stars as suggested by radial velocity surveys.  
  If gas disks disappear much faster around high-mass stars than around solar and subsolar-mass stars, 
then inward migration is halted and the planets are stranded at large semimajor axes.  
The arguments described here build on work by \citet{Bi07} who studied how a stellar-mass-dependent 
gas disk lifetime may explain why the period valley is more pronounced for planets orbiting F stars than G and K stars.
  In Section 2, I make simple analytical arguments to show that the observed 
semimajor axis distribution of giant planets may emerge from a stellar-mass-dependent disk lifetime.
 In Section 3, I perform numerical modeling similar to recent work \citep[e.g.][]{Il04, Bi07} 
to test my hypothesis.  I construct synthetic distributions of giant planets using a 
suite of Monte Carlo simulations with a range of stellar mass-dependent disk lifetimes and as well as a 
mass-independent lifetime.  A mass-independent lifetime poorly reproduces the observed 
semimajor axis distribution of planets, while a mass-dependent lifetime reproduces observed trends.
\section{Analytical Motivation}
Planets form in disks around young stars.
Once planets grow to Jovian masses, they can open a gap in the disk and undergo 'Type II migration' \citep[][]{LinPap85, Ward97} 
with a migration rate regulated by the local viscous diffusion time.  For regions interior to $\approx$ 20 AU,
migration is inward.  The migration rate can be parameterized assuming 
a Minimum Mass Solar Nebula (MMSN) model as prescribed in \citet[]{Il04} scaled to the star's mass where the initial 
gas column density is $\Sigma_{g, 1 M_{\odot}, MMSN}$ = 2400 g cm$^{-2}$ at 1 AU:
\begin{equation}
\frac{da}{dt}\sim1.3\times10^{-5} AU/yr\times(\frac{a_{p}}{1 AU})^{0.5}(\frac{M_{J}}{M_{p}})(\frac{\alpha}{10^{-3}})(\frac{M_{\star}}{M_{\odot}})^{1.5}e^{-4t/\tau_{g}}.
\end{equation}
In this equation, M$_{p}$/M$_{J}$ is the planet mass in Jovian masses, $\alpha$ is the viscosity parameter, M$_{\star}$ is the stellar mass, 
t is time, and $\tau_{g}$ is the timescale for the local disk surface density to drop to $\lesssim$ 1--2\% of 
its original value\footnote{I set the parameter f$_{g}$ in \citet{Il04} equal to M$_{\star}$/M$_{\odot}$.  Equation 
65 in \citet{Il04} is missing a factor of $\sqrt{M_{\odot}/M_{\star}}$.}.

The maximum Type II migration rate is the radial velocity of the gas, da$_{p}$/dt $\sim$ -1.5$\alpha(\frac{H}{a_{p}})^{2}a_{p}\Omega$, 
where H is the disk scale height at a$_{p}$ and $\Omega$ is the Keplerian frequency.  For a MMSN surface density profile this 
rate is
\begin{equation}
\frac{da_{p}}{dt}\sim 2.3\times10^{-5}AU yr^{-1}(\frac{M_{\star}}{M_{\odot}})^{0.5}(\frac{\alpha}{10^{-3}}).
\end{equation}
This equation governs the migration rate for disk masses that are much larger than the planet's mass.
The nominal migration timescale from Equation 1, a$_{p}$/(da$_{p}$/dt), is then $\tau_{m,II}$ $\approx$ 0.16 Myr 
for a Jovian mass planet at 1 AU with $\alpha$ = 5$\times$10$^{-4}$ if 
$\tau$$_{m,II}$ $<<$ $\tau_{g}$.  The timescale from the maximum drift rate (Equation 2) is about half this value.
If the local surface density in a gaseous disk is drained to $\lesssim$ 1\% 
of its initial value (t $\approx$ $\tau_{g}$), Equation 1 implies a migration rate of $\lesssim$ 1 AU/20Myr.  Because 
 nearly all gaseous circumstellar disks disappear by $\approx$ 10 Myr \citep{Cu07a,Cu07c}, 
migration is essentially halted.  Thus, migration is halted if 
$\tau_{g}$ $\lesssim$ $\tau_{m,II}$.

Simple arguments show that the gas disk lifetime may depend on stellar mass ($\tau_{g}$ $\propto$ M$_{\star}$$^{-\beta}$) 
and that the ratio of $\tau_{g}$ to $\tau_{m,II}$  may \textit{strongly} depend on stellar mass \citep[see also][]{Bi07}.
Optical spectroscopic studies of 1--10 Myr-old star-forming regions reveals a strong trend between accretion rate 
and stellar mass, dM$_{\star}$/dt $\propto$ M$_{\star}$$^{2}$, for 0.03--3 M$_{\odot}$-mass stars \citep{Calvet04, Muzerolle05}.
The timescale for a star to accrete some fraction of its mass, x, is then $\tau_{g}$ $\approx$ 
3Myr$\times$$(\frac{x}{0.1})(\frac{M_{\star}}{M_{\odot}})^{-1}$ or $\tau_{g}$ $\propto$ M$_{\star}$$^{-1}$.  
Thus, gas disks may dissipate faster around high-mass stars than around low-mass stars.

A shorter gas dissipation timescale eventually leads to a slower migration rate.
According to Equation 1, at t/$\tau_{g, 1M_{\odot}}$=0.2, a planet orbiting a 3 M$_{\odot}$ 
star migrates inward at a slower rate than one orbiting a 1 M$_{\odot}$ star if $\beta$ $\approx$ 1.  
At slightly later times (e.g., t/$\tau_{g, 1 M_{\odot}}$ = 0.3), the migration rate for a planet orbiting 3 M$_{\odot}$ 
star is half that for one orbiting a 1 M$_{\odot}$ star.  Therefore, migration around 
higher-mass stars decelerates earlier; $\tau_{g}$/$\tau_{m,II}$ will be smaller for higher-mass stars.  
Because smaller timescale ratios will strand more gas giants at larger semimajor axes (e.g. $\sim$ 1 AU),
 the relative frequency of hot Jupiters and other close-in planets should be lower for higher-mass stars.
\section{Numerical Model}
To test whether a stellar-mass-dependent gas disk lifetime can explain the observed 
distribution of extrasolar gas giant planets, I produce synthetic populations of exoplanets in 
semimajor axis vs. stellar mass space using a suite of Monte Carlo simulations.  
My model generally follows the approach of \citet{Bi07} 
who show simulations for a 1 M$_{\odot}$ star and vary the range in disk lifetimes (1--10 Myr, 
3--30 Myr, 10-100 Myr) to show how a stellar mass-dependent disk lifetime regulates the distribution of exoplanets.   
Here, I perform simulations for a range of stellar masses and include an explicit power law dependence 
for the gas disk lifetime.

Stellar masses for the synthetic population are randomized between 0.3 and 3 M$_{\odot}$ with 
a probability distribution weighted towards solar/subsolar-mass stars (P(M$_{\star}$) $\propto$ M$_{\star}$$^{-2.5}$: 
a Salpeter-like IMF).  To model the regions of planet formation,
 the planets' initial semimajor axes are chosen between 1 and 25 AU.  For my fiducial model, 
I assume a gas disk lifetime that scales inversely with stellar mass ($\tau_{g}$ = 4 Myr$\times$(M$_{\star}$/M$_{\odot}$)$^{-1}$).  

 For Jovian-mass planets to form, I require that they reach an isolation core mass 
of M$_{iso}$=5 M$_{\oplus}$ \citep[e.g.][]{Alibert2005}.  I assume a 2.5$\times$ scaled MMSN model from 
\citet{Il04} to account for the median disk mass needed to form cores \citep{Kb08} and set the metallicity comparable to the 
median metallicity of stars with detected Jovian-mass planets ([Fe/H] $\sim$ 0.15, \citealt{Wright09}).
  I set the ice line location equal to values from Figure 1 of \citet[][]{Kk08} at 0.3 Myr 
interpolating between values for stars of different masses\footnote{\citet{Il04} and 
\citet{Bi07} assume an optically-thin disk in determining the ice line position.
  Like \citet{Kk08}, I assume that the disk is not optically thin when planetesimals grow into gas giant cores.}.
Finally, I require that enough gas is left when the core mass is 5 M$_{\oplus}$
 to form a Jovian-mass planet.  This condition is equivalent to 
\begin{equation}
M_{g, iso} = 1.34 M_{J}\times(\frac{a_{p}}{1AU})^{0.75}(\frac{\Sigma_{1AU, 1M_{\odot}}}{2400 g cm^{-2}})^{1.5}(\frac{M_{\star}}{M_{\odot}}),
\end{equation}
where the gas feeding zone size is 10 Hill radii.  The Jovian planet-forming regions are 
$>$ 2 AU for 0.5 M$_{\odot}$ stars, $>$ 3 AU for 1 M$_{\odot}$ stars, and $>$ 5 AU for 2.5 M$_{\odot}$ stars.

From their birthplaces, I track the semimajor axis evolution of planets from Type II migration 
according to Equations 1 and 2, assuming a viscosity parameter of $\alpha$=5$\times$10$^{-4}$.
Planets reaching within $\approx$ 2--5 R$_{\star}$ ($\sim$ 2--5 R$_{\odot}$(M$_{\star}$/M$_{\odot}$)$^{0.75}$ 
 may be affected by magnetospheric disk truncation \citep{Lin96}, which is not treated in this simple model.  
Any planet that reaches this small separation is given a final semimajor axis that is randomized between 2 and 5 R$_{\star}$.
I make two simplifying assumptions in the fiducial model that will be removed later.  First, I assume that all planets are the 
mass of Jupiter.  Second, I assume that all planets form at 1 Myr: a characteristic time for the formation of 
5 M$_{\oplus}$ cores at 5 AU around solar-mass stars \citep{Kb08}.

\begin{figure}
\epsscale{0.9}
\centering
\plottwo{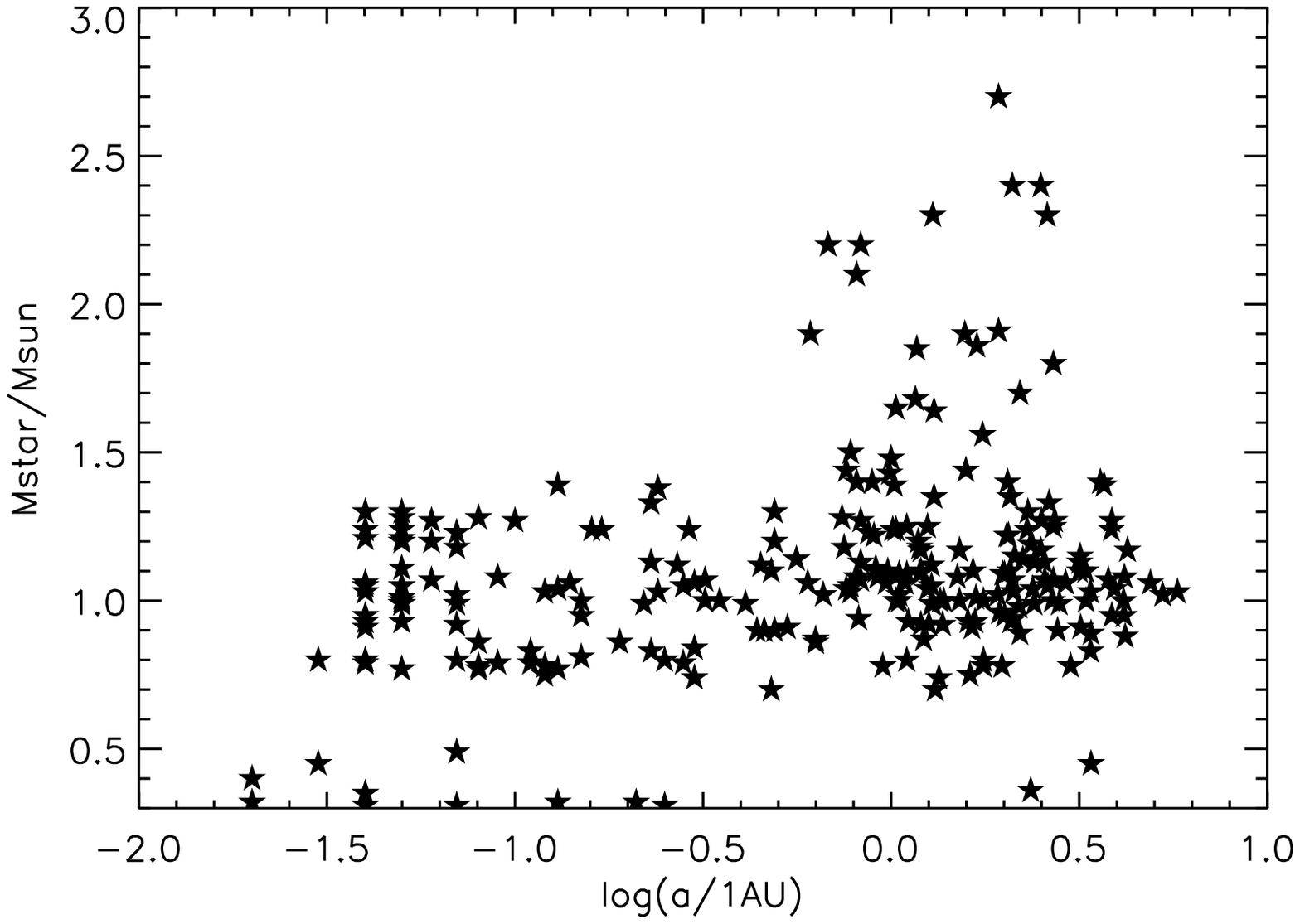}{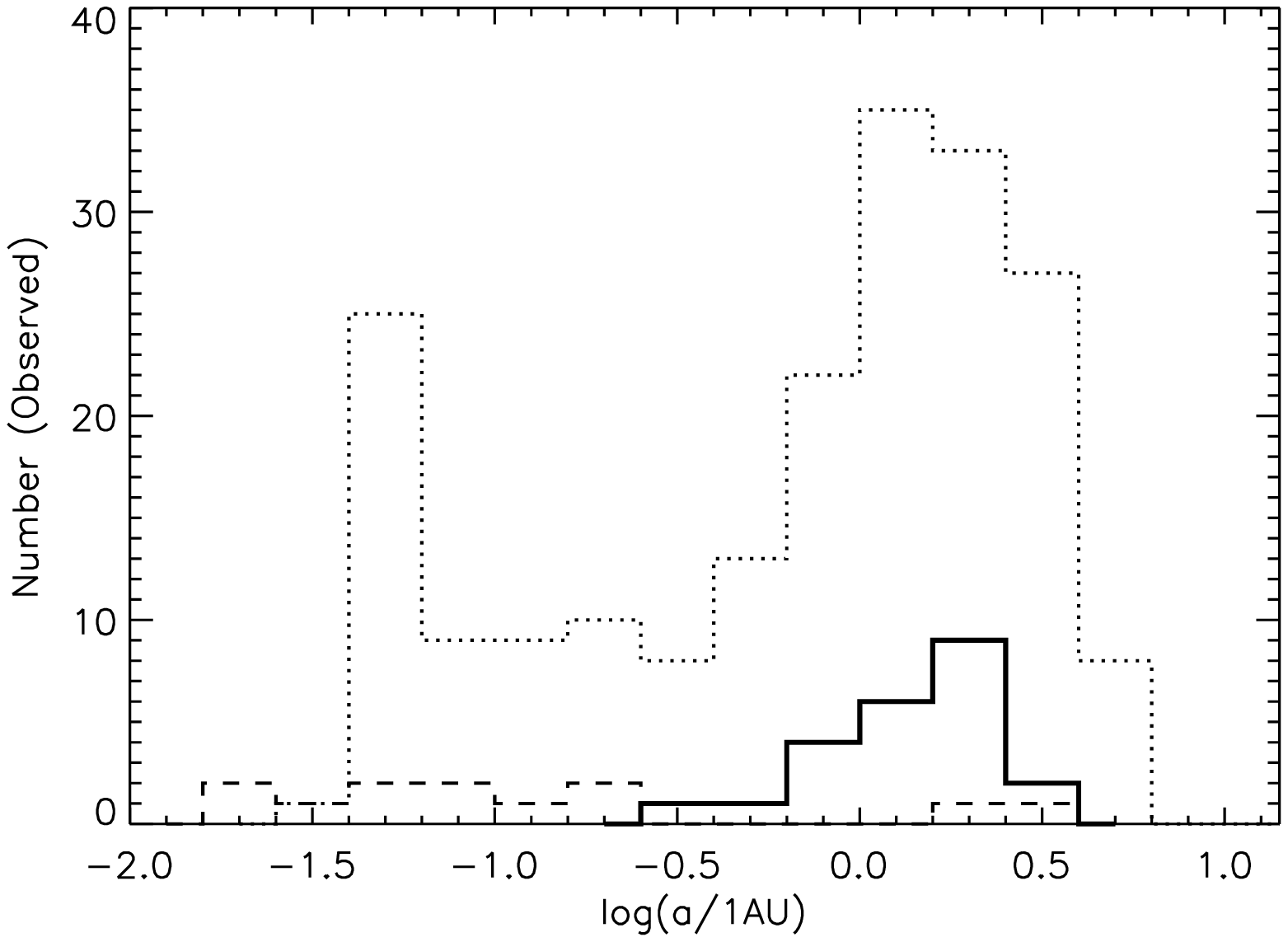}
\plottwo{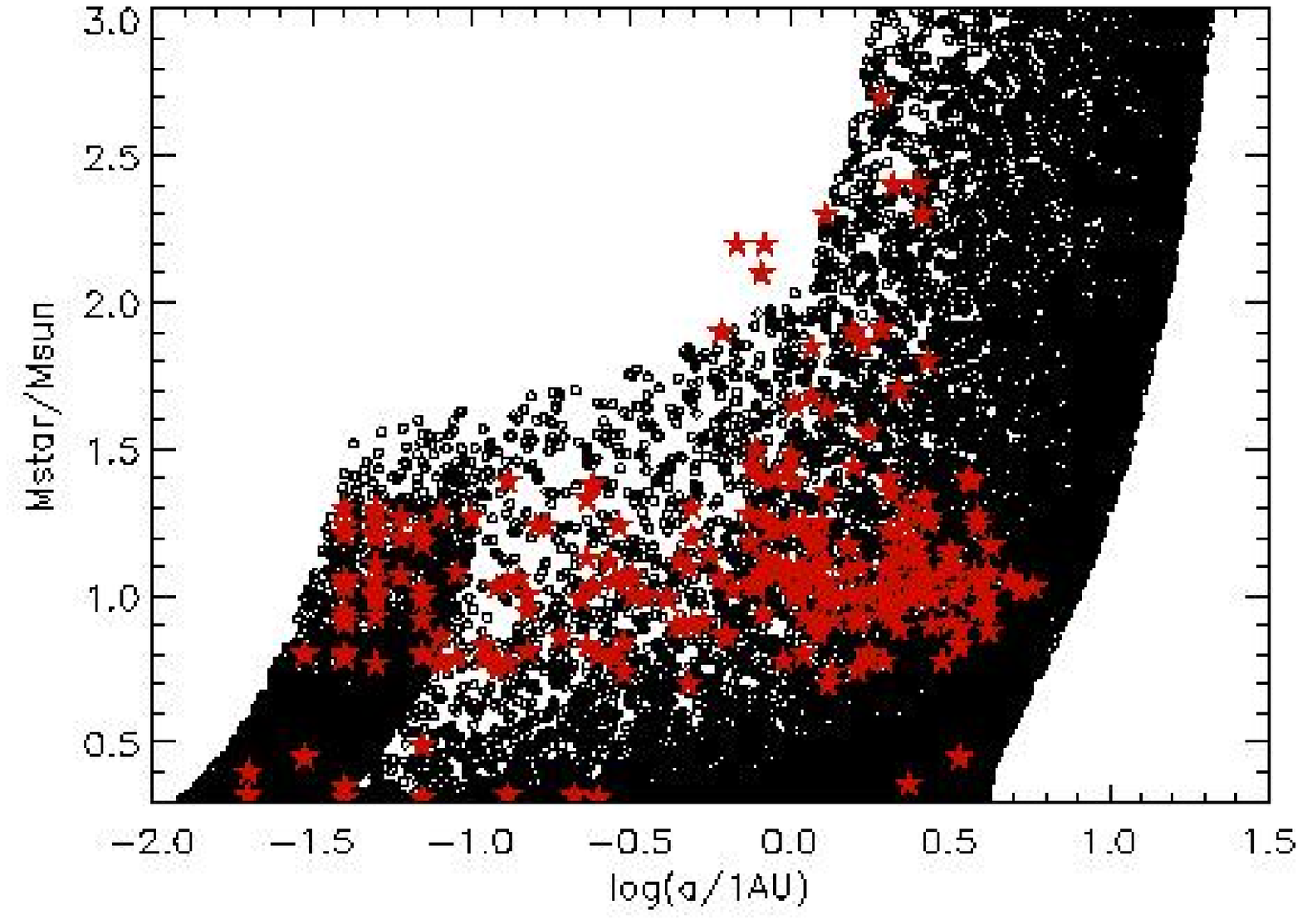}{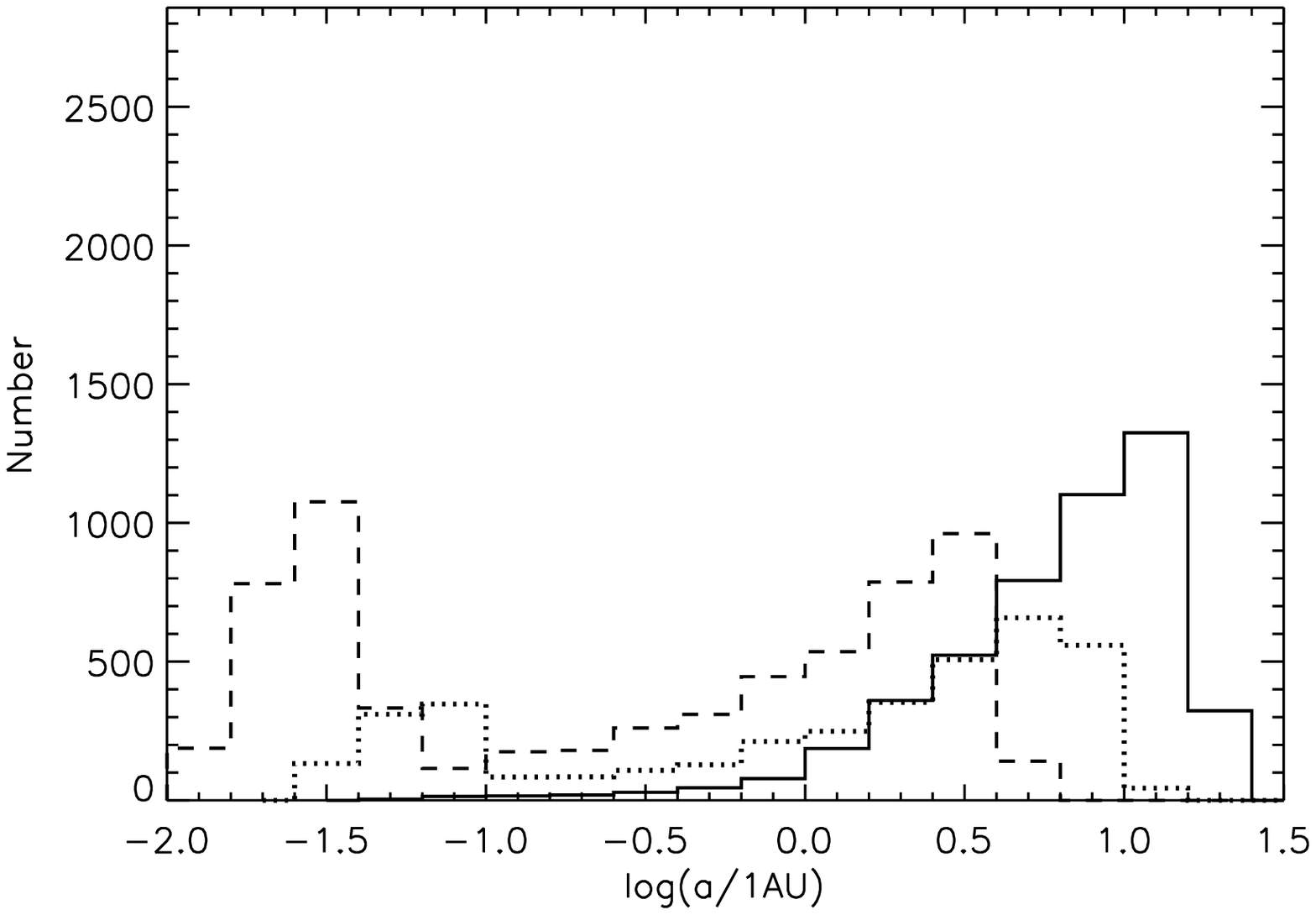}
\caption{Top panels: the observed semimajor axis vs. stellar mass distribution  
 (left) and the histogram plot of semimajor axes from radial velocity surveys.
Bottom panels: the semimajor axis vs. stellar mass distribution for the
synthetic population (black circles) with my fiducial model, assuming $\tau_{g}$ $\propto$ M$_{\star}$$^{-1}$ (left) 
and distribution of radial velocity-detected planets (red stars).  (right) Histogram 
plots of the semimajor axis distribution for 0.3--0.5 M$_{\odot}$ stars (dashed line), 0.8--1.5 M$_{\odot}$ 
stars (dotted line), and 1.5--3 M$_{\odot}$ stars (solid line).}
\label{fg1}
\end{figure}

Table 1 lists simulation results for a total of 20,000 planets (20 simulations of 
1000 planets; N$_{total}$).  The table shows the total number of planets with final semimajor 
axes $\le$ 3 AU and in three semimajor axis bins ($<$ 0.2 AU, 0.2--0.5 AU, and 0.5--3 AU) for 
each of the stellar mass bins (0.3--0.5 M$_{\odot}$, 0.8--1.5 M$_{\odot}$, and 1.5--3 M$_{\odot}$).
The lower left panel of Figure \ref{fg1} shows the final semimajor axis versus stellar mass distribution of these planets (black circles).  
I overplot the distribution of radial velocity-detected planets with 
well-constrained stellar masses (258; red stars\footnote{Downloaded from http://exoplanet.eu.}). 
The lower right panel shows a histogram plot of the semimajor axes for the synthetic population in three mass bins: 
0.3--0.5 M$_{\odot}$ (dashed line), 0.8--1.5 M$_{\odot}$ (dotted line) and 1.5--3 M$_{\odot}$ (solid line).  

Despite its simplicity, the model yields good agreement with observed semimajor axis distributions.
  In the synthetic population, stars with masses between 
0.3 M$_{\odot}$ and $\sim$ 1 M$_{\odot}$ have many hot Jupiters.  The right panel of Figure \ref{fg1} shows 
that few 0.8--1.5 M$_{\odot}$ stars have planets with a$_{p}$ $\sim$ 
0.1 and 0.5 AU, consistent with the 'period valley' in the observed population.  

Most strikingly, the synthetic population shows a sharp drop in the number of hot Jupiters from 
$\sim$ 1.3 M$_{\odot}$ to 1.5 M$_{\odot}$.  For my model assumptions, the synthetic population lacks any hot Jupiters
 for stars with M$_{\star}$ $\gtrsim$ 1.6 M$_{\odot}$ and lacks any planets with intermediate (0.2--0.5 AU) distances for 
M$_{\star}$ $\gtrsim$ 1.8 M$_{\odot}$.  This trend is consistent with the clear lack of detected hot Jupiters 
around 1.5--3 M$_{\odot}$ stars.  The $\chi^{2}$ values comparing the relative frequencies 
of hot Jupiters, planets at 0.2--0.5 AU, and those at 0.5--3 AU with the observed distribution confirm 
that agreement is good, especially for $\gtrsim$ 0.8 M$_{\odot}$ stars.

\begin{figure}
\centering
\epsscale{0.99}
\plottwo{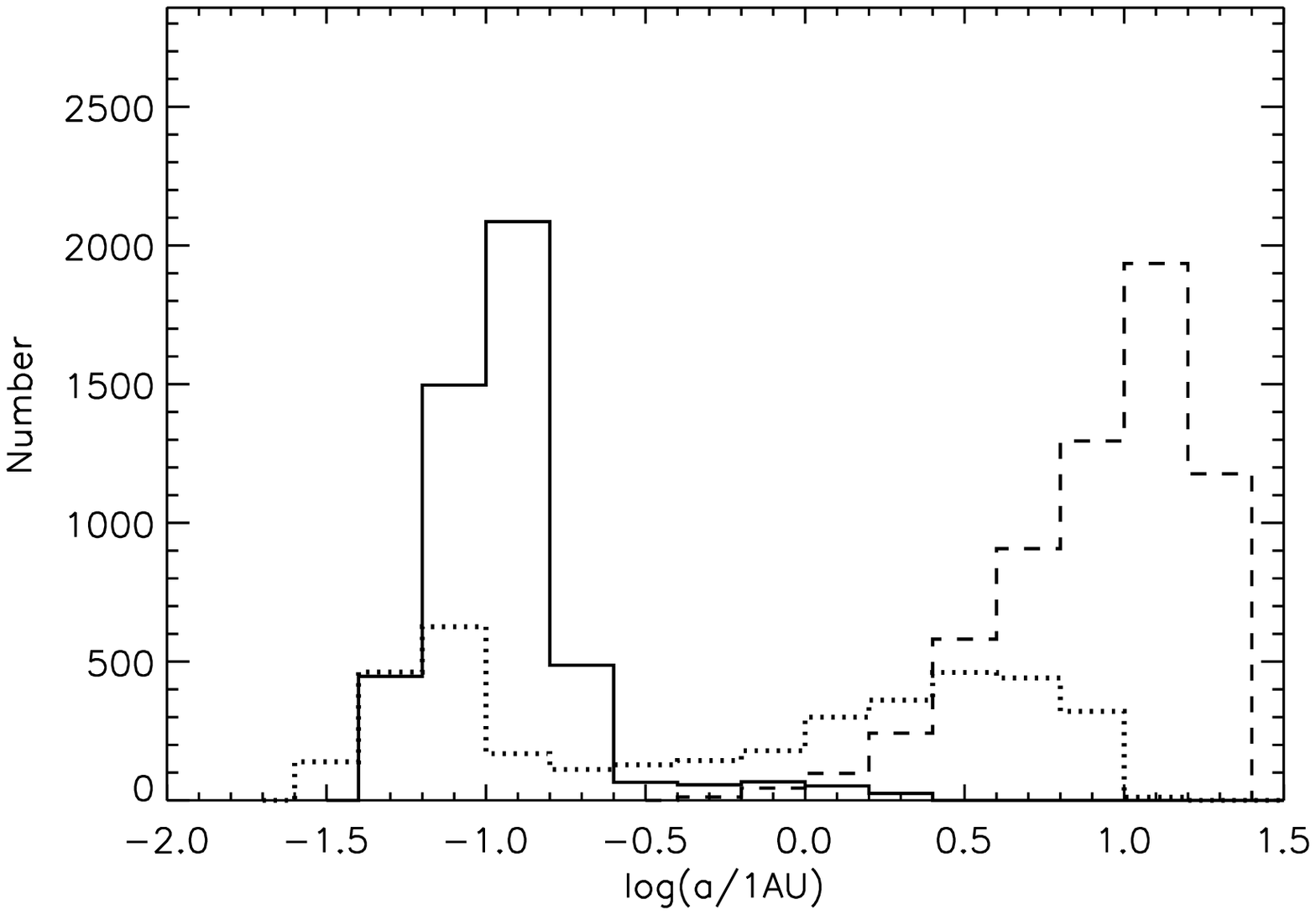}{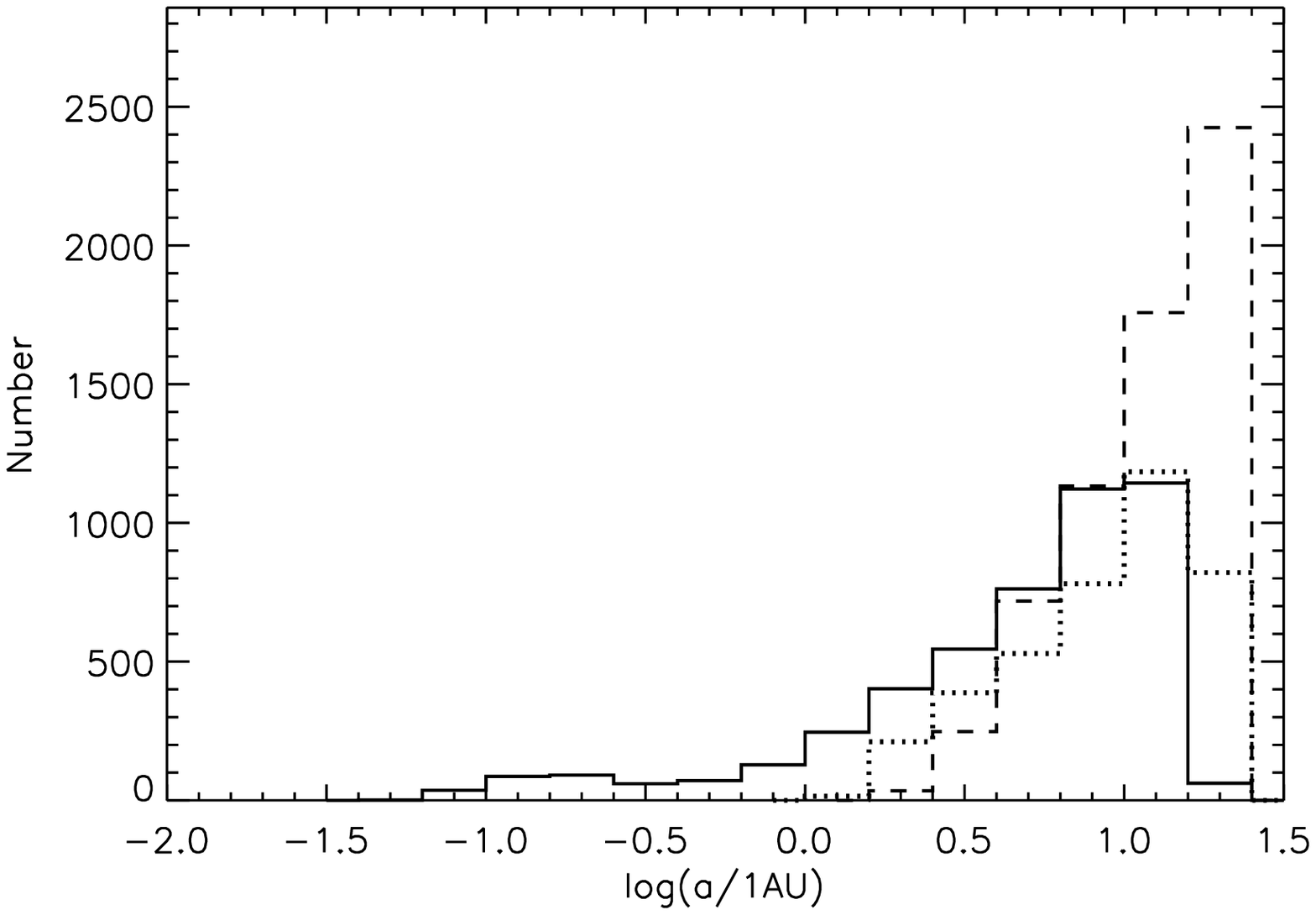}
\plottwo{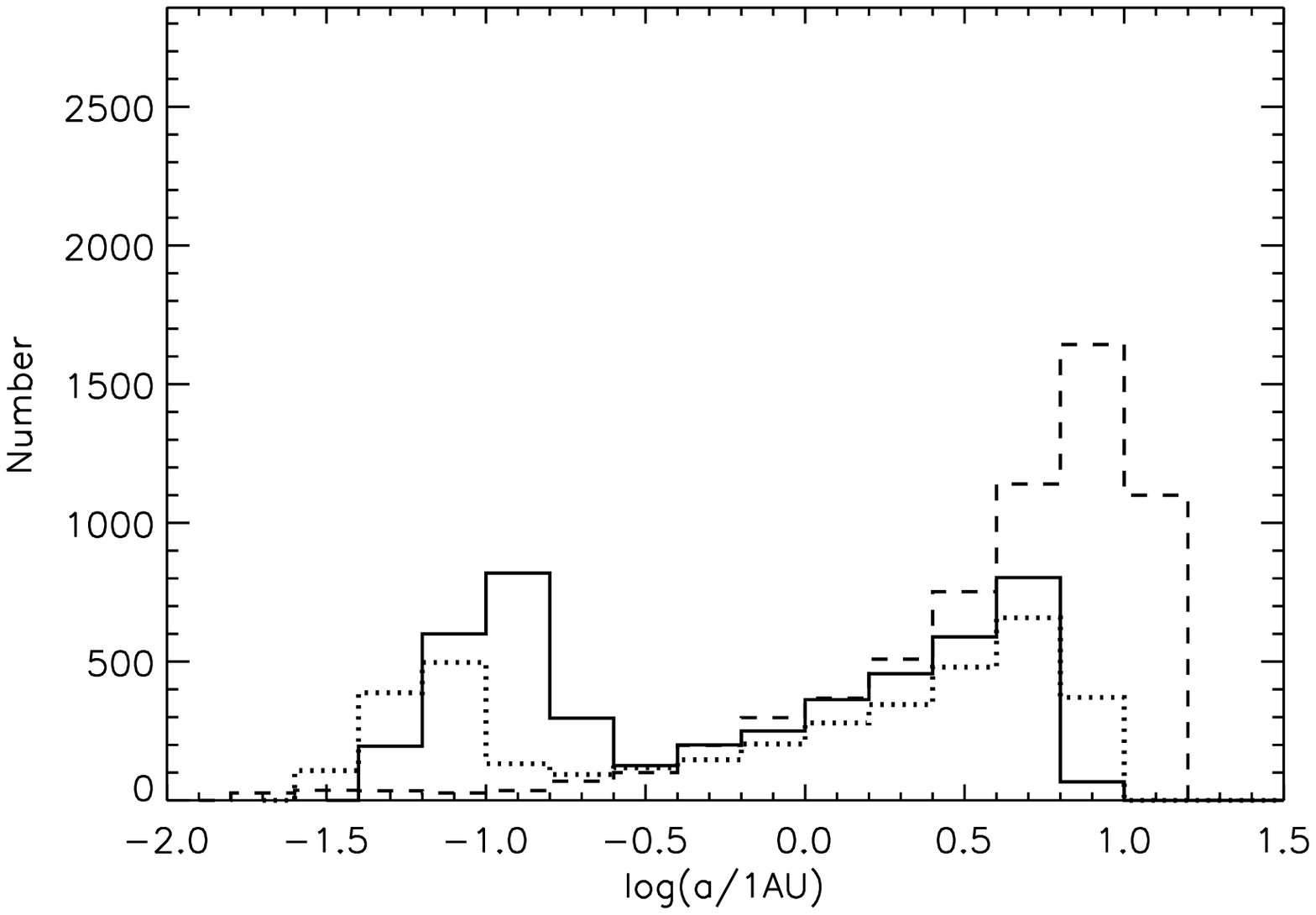}{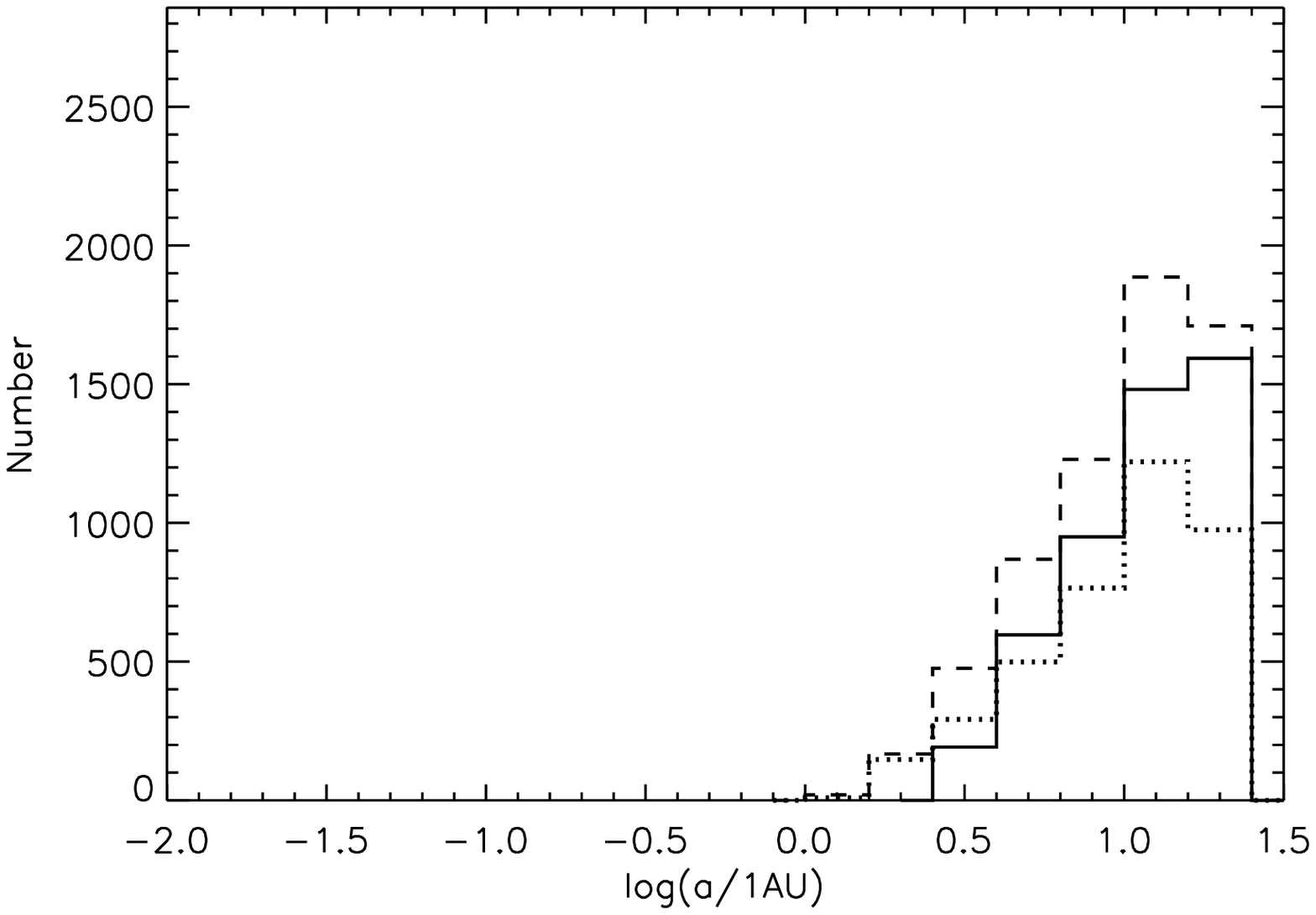}
\caption{Histogram plots of the semimajor axis distribution assuming $\tau_{g}$= 4Myr (top left), 
$\tau_{g}$= 2 Myr (top right), $\tau_{g}$ = 4 Myr$\times$(M$_{\star}$/M$_{\odot}$)$^{-0.5}$ (bottom left), 
and $\tau_{g}$ = 2 Myr$\times$(M$_{\star}$/M$_{\odot}$)$^{-0.5}$ (bottom right) for my fiducial model.  
  Each distribution is inconsistent with the observed 
population.}
\label{fg2}
\end{figure}

The semimajor axis distribution shows far poorer agreement if the gas disk lifetime is independent of or 
weakly dependent on stellar mass (Figure \ref{fg2}).   Assuming $\tau_{g}$=4 Myr for all masses turns all planets 
around $>$ 1.5 M$_{\odot}$ stars into hot Jupiters, weakens the period valley for solar-mass stars, and 
confines all planets around low-mass stars to semimajor axes with a$_{p}$ $>$ 10 AU (top-left panel).  These 
properties are inconsistent with the observed distribution.  
The model with $\tau_{g}$ = 2 Myr (top-right panel) for all stars yields the correct distribution for high-mass 
stars but eliminates all hot Jupiters around solar/subsolar-mass stars, and strands all planets around subsolar-mass 
stars at a$_{p}$ $>$ 10 AU.  Models with $\tau_{g}$ $\propto$ M$_{\star}$$^{-0.5}$ fare marginally better (bottom panels).  
The model with $\tau_{g, 1 M_{\odot}}$ = 4 Myr predicts a pileup of hot Jupiters, many planets at $\sim$ 3--10 AU, 
and a dearth of planets at $\sim$ 0.2--0.5 AU for all stars with masses $>$ 0.8 M$_{\odot}$.  
It also confines planets around $<$ 0.5 M$_{\odot}$ stars to $>$ 0.3 AU with a peak 
at $\sim$ 10 AU.  The model with $\tau_{g, 1 M_{\odot}}$ = 2 Myr confines all planets to a$_{p}$ $>$ 3 AU.
These features are clearly not present in the observed population.
The $\chi^{2}$ values for all of these simulation runs exceed 100 for low-mass stars and (sometimes) 
high-mass stars.

Motivated by the success of the fiducial model with $\tau_{g}$ $\propto$ M$_{\star}$$^{-1}$, 
I remove assumptions regarding the planet's mass and core formation timescale and run a second set of simulations. 
First, I set the planet's mass equal to the minimum gap-opening mass for Type II migration, requiring that the planet's Hill radius 
is larger than the local disk scale height \citep[e.g.][]{LinPap85}:
\begin{equation}
M_{II} \sim 0.4 M_{J}\times(\frac{M_{\star}}{M_{\odot}})(\frac{a_{p}}{1 AU})^{0.75}.
\end{equation}
Second, I estimate the formation timescale for each core explicitly, require that the core can form before gas is dissipated, 
and require that enough gas is left to form a migrating planet after core formation.  From the \citet{Kb08} results, I extrapolate 
the formation timescale of 1 Myr at 5 AU around a solar-mass star with $\Sigma_{d}$ = 2.5 g cm$^{-2}$ 
to different distances, stellar masses, and dust column densities.  I assume that 
$\tau_{core}$ $\propto$ ($\Sigma_{d}$$\Omega$)$^{-1}$, which 
is a reasonable approximation of the numerical results \citep{Kb08b}.

\begin{figure}
\epsscale{0.85}
\centering
\plottwo{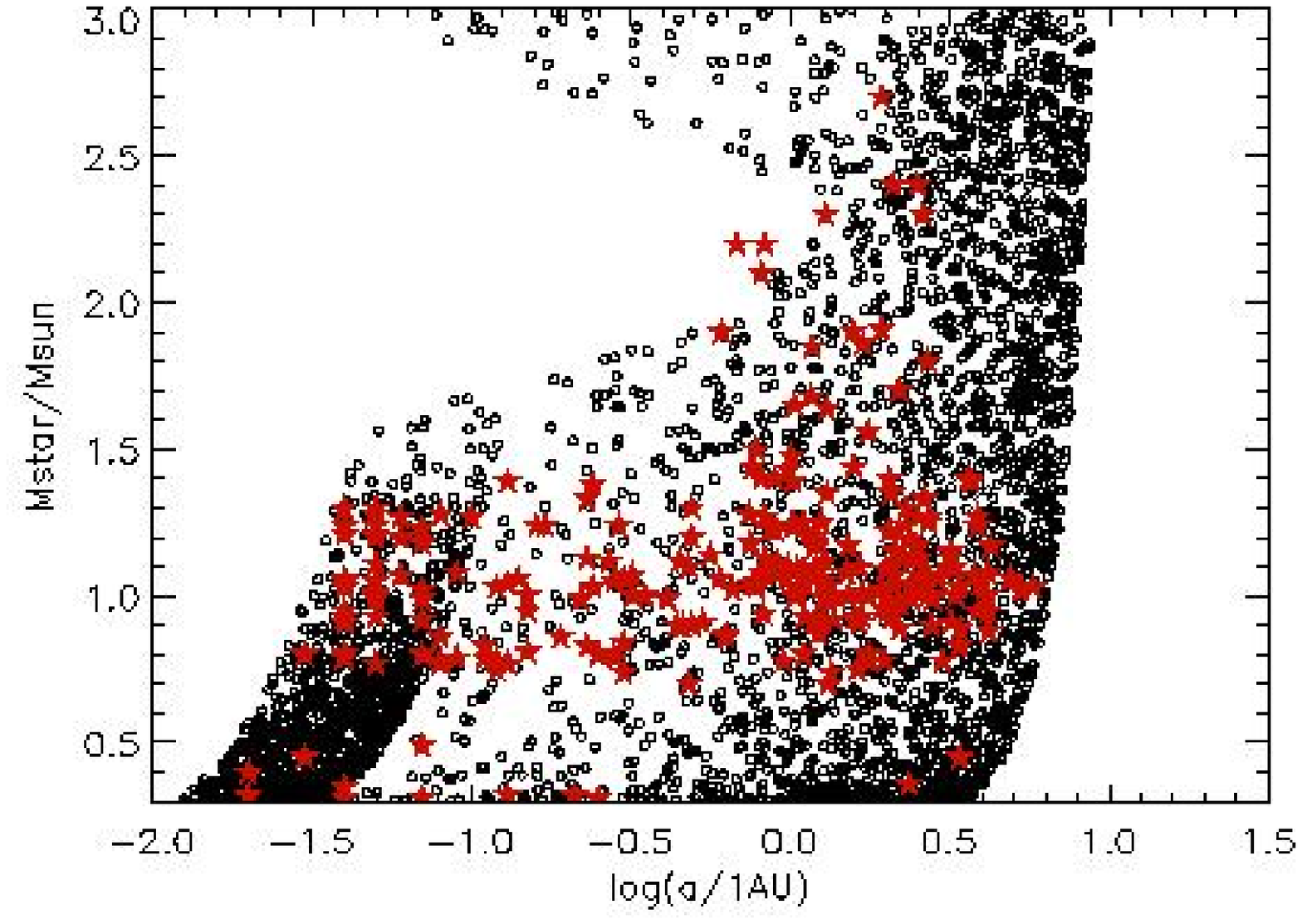}{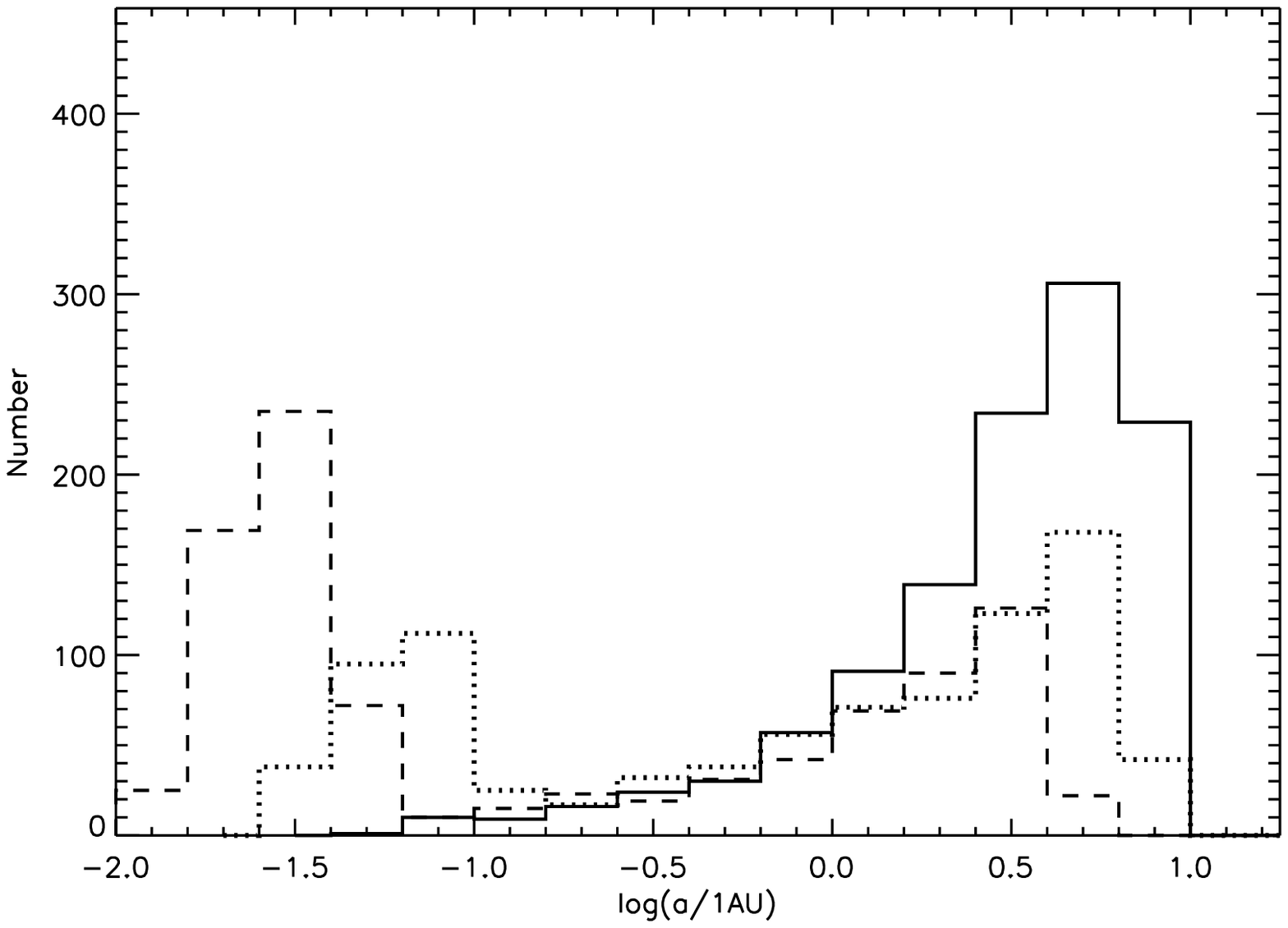}
\plottwo{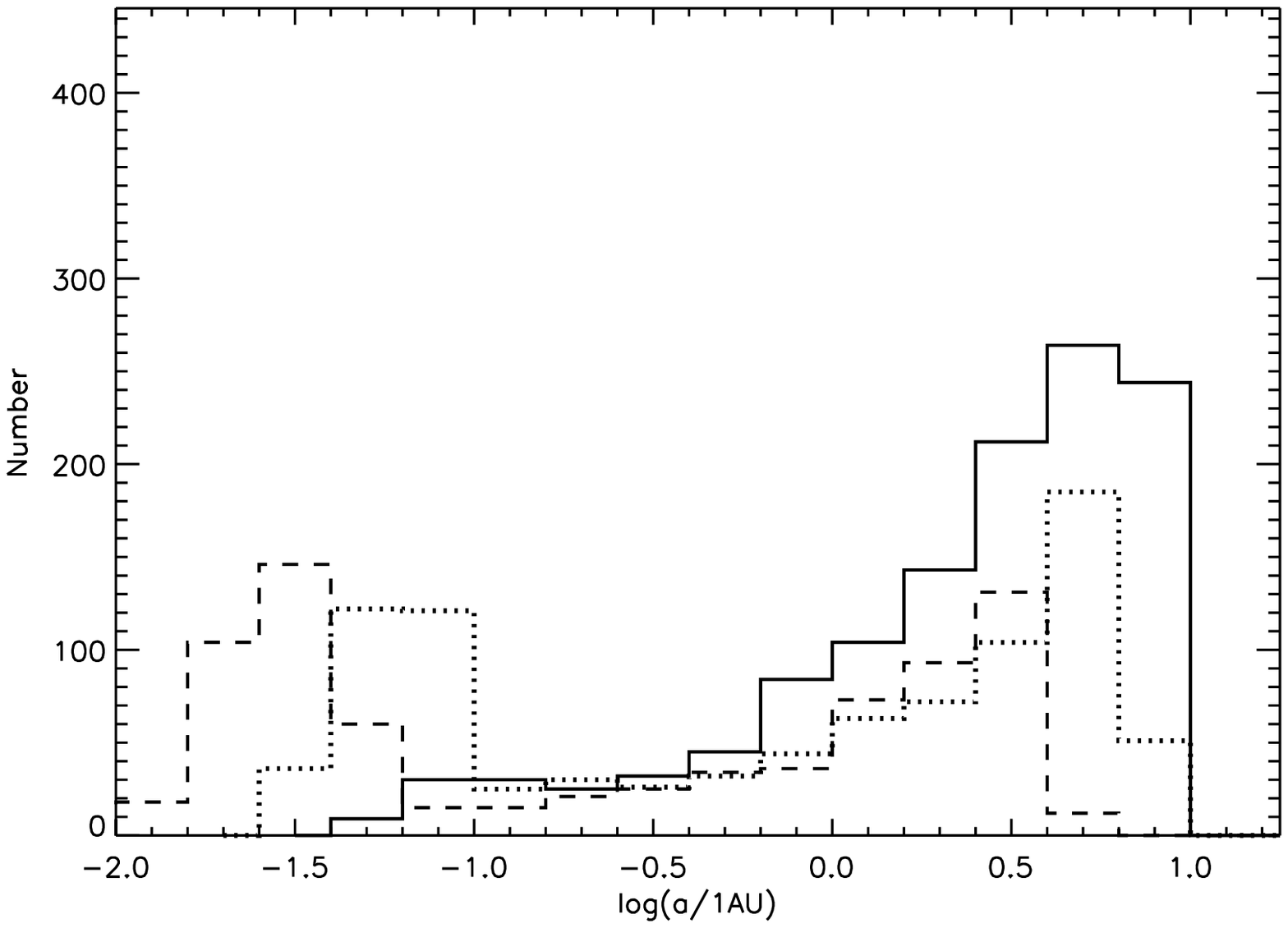}{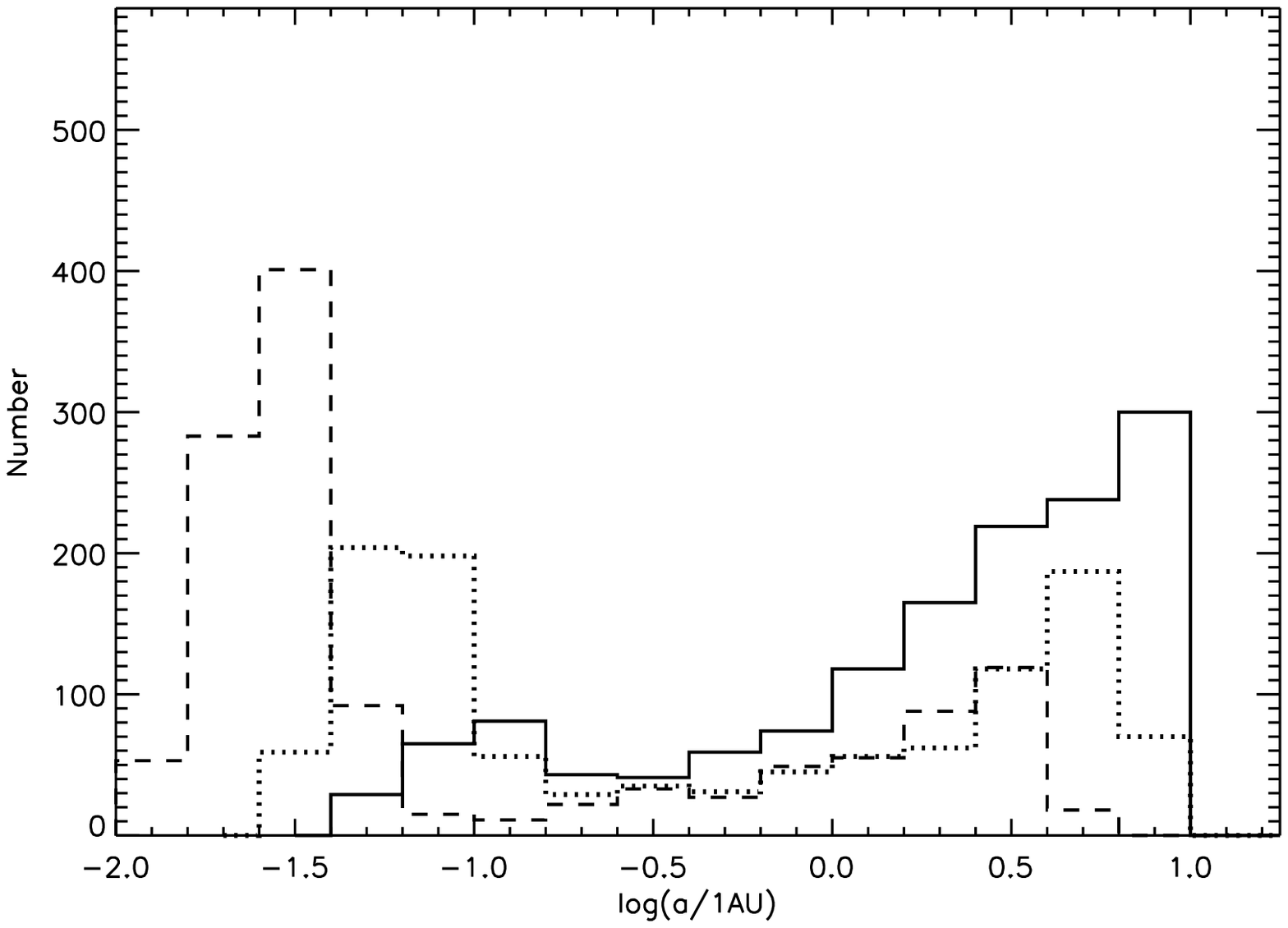}
\plottwo{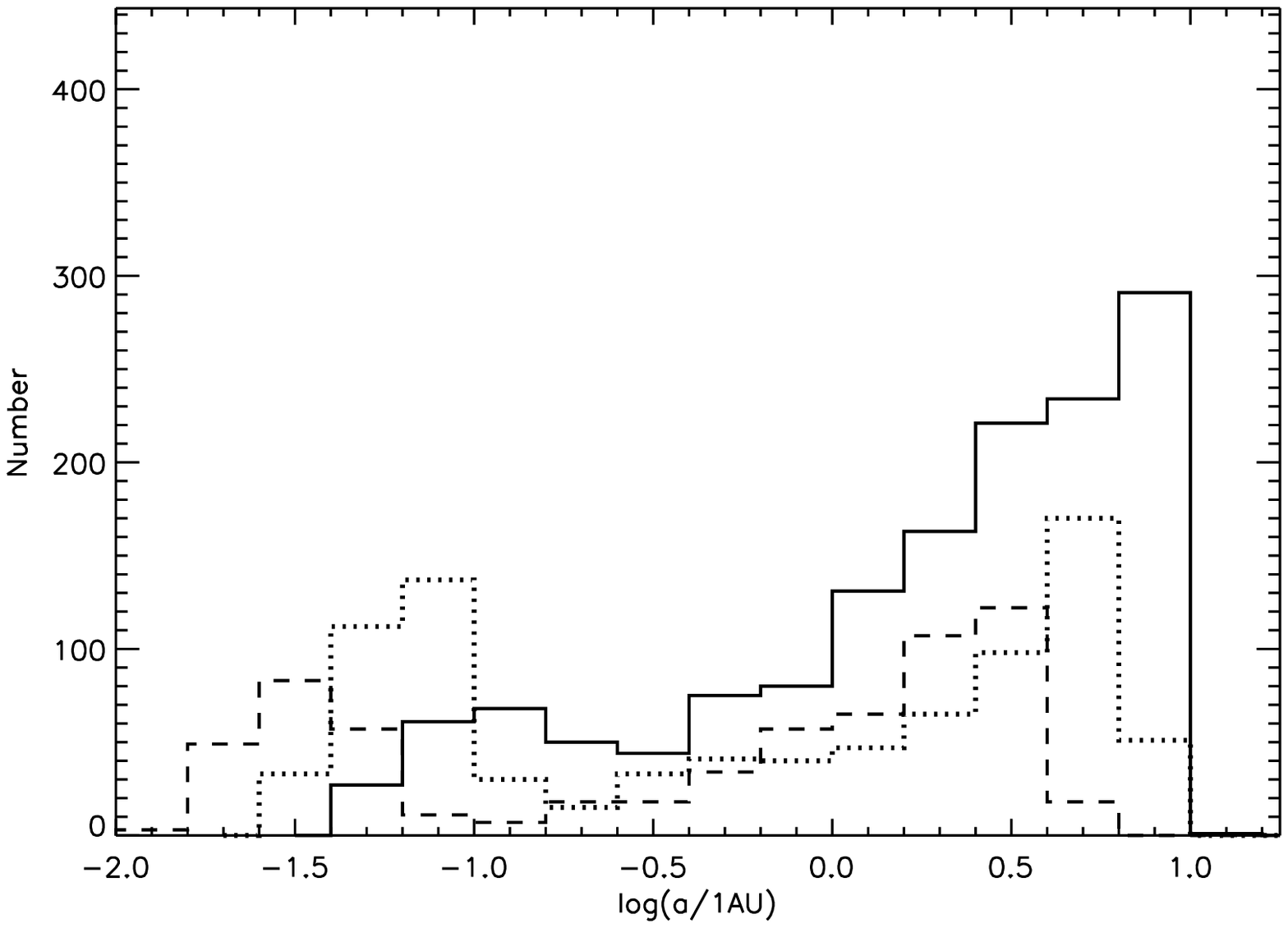}{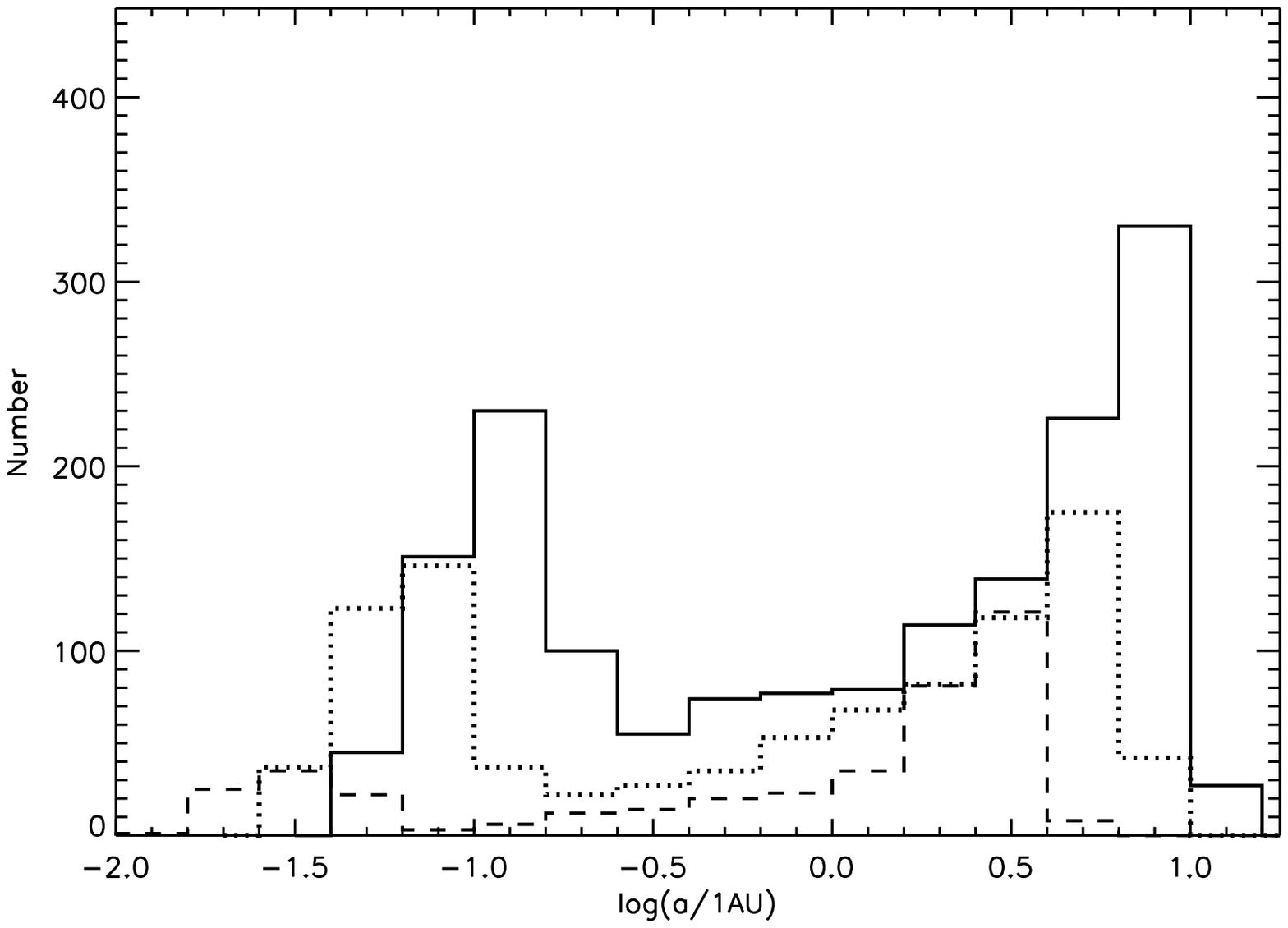}
\caption{Results for the second set of simulations.  
Top left: distribution of semimajor axis vs. stellar mass and histogram plot 
of the semimajor axis distribution (top right) for  
$\tau_{g}$ = 2 Myr$\times$ (M$_{\star}$/M$_{\odot}$)$^{-1}$.  
Second row: histogram plots for $\tau_{g}$ = 2 Myr$\times$ (M$_{\star}$/M$_{\odot}$)$^{0.75}$ (left)
and $\tau_{g}$=3 Myr$\times$(M$_{\star}$/M$_{\odot}$)$^{-1}$ (right).
Bottom row: histogram plots for $\tau_{g}$= 2 Myr$\times$ (M$_{\star}$/M$_{\odot}$)$^{-0.5}$ (left) and 
$\tau_{g}$= 2 Myr$\times$ (M$_{\star}$/M$_{\odot}$)$^{-0.1}$ (right).}
\label{fg3}
\end{figure}

Results from the second set of simulations (the bottom half of Table 1, Figure \ref{fg3}) show good agreement 
between the synthetic distribution and the observed distribution if $\tau_{g}$ $\propto$ M$_{\star}$$^{-\beta}$, 
where $\beta$ = 0.75-1.5.  The semimajor axis versus stellar mass distribution (shown for $\tau_{g}$=2 Myr$\times$ 
M$_{\star}$$^{-1}$, top-left panel) features the same sharp drop in the frequency of hot Jupiters and planets 
with a$_{p}$ = 0.2--0.5 AU (log(a$_{p}$)= -0.7-- -0.3) for M$_{\star}$ $\ge$ 1.5 M$_{\odot}$ 
exhibited by the fiducial model.  The histogram plots for $\beta$ = 1 (top right panel) and $\beta$ = 0.75 (second row, 
left panel) confirm that there is a lack of hot Jupiters and other planets with a$_{p}$ $<$ 0.5 AU for high-mass stars.
The models successfully reproduce the two other features of the observed distribution: 
the period valley for solar-mass stars and the lack of long-period planets around 
low-mass stars.  The agreement is confirmed quantitatively as models with $\beta$ = 0.75--1.5 typically have 
$\chi^{2}$ values less than $\sim$ 20 for all stellar mass bins.   

I also calculate the percentage of planets that achieve a gap-opening mass and undergo Type II migration (N$_{gap}$)
for each of the stellar mass bins (100$\times$N$_{gap}$/N$_{total}$).  
For $\beta$ $<$ 1.5, the frequency is lowest for M stars.  
This percentage probes the relative sizes of planet-forming 
regions for a given scaled disk mass (2.5 $\times$ a scaled MMSN) and is different from but related to 
the frequency of planets \citep[][]{Jj07b}. 
For calculations with lower scaled disk masses (e.g., 1.25$\times$ MMSN scaled) the planet-forming regions shrink, 
and only solar-to-high mass stars form migrating planets.  For calculations with an even lower 
scaled disk mass, only high-mass stars form migrating planets.   The percentage of 
migrating gas giants for M stars is lowest because 
their disks have lower masses: M$_{disk}$ $\propto$ f$_{g}$ $\propto$ M$_{\star}$ \citep{Kk08}.

Models with $\tau_{o}$ = 2 Myr and $\beta$ $\le$ 0.5 (third row) begin to show disagreement with the observed distribution as they 
predict a pileup of hot Jupiters and period valley for high-mass stars and a lack 
of hot Jupiters for low-mass stars.  The $\chi^{2}$ values for low-mass stars and high-mass stars for $\beta$ $\le$ 0.5 
are significantly higher than for $\beta$ = 0.75--1.5.  Therefore, according to my simulations, 
the observed semimajor axis distribution of gas giant planets results from a stellar-mass-dependent gas disk lifetime.
\section{Discussion}
Through a series of Monte Carlo simulations, I have shown that a stellar-mass-dependent gas disk lifetime can  
explain the observed semimajor axis distribution of extrasolar gas giant planets, including the lack of hot Jupiters 
and other planets with a$_{p}$ $<$ 0.5 AU around high-mass stars.  Synthetic distributions of planets 
produced assuming that the gas disk lifetime, $\tau_{g}$, scales as M$_{\star}$$^{-\beta}$ with $\beta$ = 0.75--1.5 
 reproduce key features in the observed semimajor axis distribution.   Distributions from models lacking a 
stellar-mass-dependent disk lifetime quantitatively provide a poorer match to observations.

This work extends and complements the investigation of \citet{Bi07} who use Monte Carlo simulations 
and semi-analytical prescriptions for planet growth to explain exoplanet trends for solar-mass stars.
While high-mass stars lack a pronounced period valley, predicted 
by \citet{Bi07}, the dearth of planets at 0.2--0.5 AU agrees with 
their predictions.  Moreover, this work shows that a stellar mass-dependent disk lifetime, invoked by \citeauthor{Bi07} 
to explain exoplanet trends for solar-mass stars, may explain trends for planets around 
stars with a wide range of masses.  Future modeling work is necessary to test this hypothesis more conclusively.  
Future modifications include incorporating a more sophisticated treatment of circumstellar gas accretion, 
modeling Type I migration, determining the sensitivity of the planets' synthetic distributions to 
 the migration rate (i.e., value of $\alpha$), and tracking the migration of planets 
\textit{while} they are accreting gas \citep[e.g.][]{Alibert2005}.

Recent studies of young stars in clusters 
support a stellar mass-dependent gas disk lifetime \citep{Kk09}.  Based on optical spectroscopy, 
the frequency of gas accretion in 2--15 Myr-old clusters is significantly higher for 
stars with M$_{\star}$ $<$ 1 M$_{\odot}$ than for higher-mass stars \citep[e.g., IC 348, Tr37, and 
h and $\chi$ Persei][]{Dahm08, Ck08, Si06, Cu07c}.  Secondary characteristics of gas-rich 
 disks (optically thick thermal infrared emission) are also rarer 
for high-mass stars \citep[e.g.,][]{Ca06, Cl09}.  
Combining cluster data to empirically constrain 
 $\tau_{g}$(M$_{\star}$) may be possible.  However, uncertainties in stellar 
ages for stars in the youngest ($<$ 3 Myr) clusters present a 
strong challenge to constructing an empirically based gas disk lifetime.
I will address these issues in a future paper.

If the gas disk lifetime strongly depends on stellar mass, my model simulations 
suggest that future radial velocity surveys will find few gas giant planets orbiting 
at small separations from high-mass stars.  If $\approx$ 10\% of 
high-mass stars have gas giant planets with a$_{p}$ $<$ 3 AU \citep{Jj07b}, the 
simulations with $\beta$ = 0.75--1.5 imply that out of 1,000 high-mass stars targeted for radial velocity 
surveys, fewer than $\approx$ 15 will have planets at a$_{p}$ $\le$ 0.5 AU 
while more than $\approx$ 75 will have planets at a$_{p}$ $\ge$ 0.5 AU.  Ongoing surveys will provide 
a larger sample from which to compare observed and predicted frequencies of hot Jupiters 
and other gas giants at small separations (J. Johnson, in preparation).
\acknowledgements
I thank John A. Johnson and Jason Wright for discussions on exoplanet data 
and for encouraging the writing of this paper; Geoff Marcy, Scott Kenyon, Dan Fabrycky, 
Ruth Murray-Clay, Brad Hansen, and Jonathan Irwin for valuable comments; and the referee for a rapid and insightful 
report.  This work is supported by NASA Astrophysics Theory Grant NAG5-13278 and NASA Grant NNG06GH25G.

\begin{deluxetable}{llllllllllllllllllllll}
\tiny
\setlength{\tabcolsep}{0.02in}
\tabletypesize{\tiny}
\tabletypesize{\scriptsize}

\tablecolumns{22}
\tablewidth{0pc}
\tabletypesize{\scriptsize}
\tablecaption{Simulation Results}
\tablehead{
\colhead{ID}&\colhead{$\tau_{o}$}&\colhead{$\beta$}&\multicolumn{3}{c}{Number}
&\multicolumn{9}{c}{Fractions of Planets in Each Semimajor Axis Bin}&\multicolumn{3}{c}{$\chi^{2}$}&\multicolumn{3}{c}{\%}\\
{}&{}&{}&{}&{}&{}&\multicolumn{3}{l}{a$_{p}$$<$0.2 AU}
&\multicolumn{3}{c}{a$_{p}$=0.2--0.5 AU}&\multicolumn{3}{l}{a$_{p}$=0.5--3 AU}\\
{}&{}&{}&{LM}&{IM}&{HM}&{LM}&{IM}&{HM}&{LM}&{IM}&{HM}&{LM}&{IM}&{HM}&{LM}&{IM}&{HM}&{LM}&{IM}&{HM}}
\startdata
Obs.&-&-&11&176&20&0.73& 0.27& 0.0& 0.18& 0.13& 0.0 &0.09& 0.60& 1.00 & --&--&-- &1.8$^{a}$& 4.2$^{a}$& 8.9$^{a}$\\
M1&4&1&5524&2198&925&0.50&0.42&0.05&0.09&0.10&0.05&0.41&0.49&0.89&28.8&6.3&3.0\\
" &4&0.5&1957&2469&3499& 0.10&0.48& 0.52& 0.11& 0.09& 0.09& 0.79& 0.43& 0.39&157.0&13.1&114.5\\
" &4&0&602&2787&4782& 0&0.52&0.94& 0& 0.09& 0.03& 0.99&0.39&0.04&245.2&19.0&317.9\\
" &2&0.5& 329& 255&20& 0&0&0&0&0&0&1.00&1.00&1.00&245.2&44.0&0\\
" &2&0& 92& 365&1290&0&0&0.14&0&0&0.09&1.00&1.00&0.77&245.2&44.0&14.2\\
M2&2&1&948& 893&1146 & 0.64& 0.46& 0.06& 0.05& 0.10& 0.10& 0.32& 0.44& 0.84&13.7&11.0&6.9&13.4&20.8&20.9\\ 
"&2&1.5&1337& 629& 234 & 0.76& 0.48& 0.0& 0.04& 0.10& 0.03 & 0.2& 0.41& 0.97 &5.7&14.3&0.3&19.6&21.4&14.5 \\
"&2&0.75& 783& 911&1222&0.53& 0.52& 0.15& 0.07& 0.10& 0.10& 0.40& 0.38& 0.75&25.1&19.7&16.7&11.1&20.7&22.4\\
"&2&0.5& 546& 584& 780& 0.40& 0.55& 0.24& 0.08& 0.09& 0.13& 0.52& 0.36& 0.63&53.5&24.2&37.2&9.2&20.4&26.1\\
"&2&0.1& 319& 671& 975 & 0.31& 0.53& 0.51& 0.09& 0.08& 0.12 & 0.60& 0.40& 0.36&78.3&19.4&120.4&5.8&22.0&30.3\\
"&3&1&1173& 810& 751& 0.74& 0.65& 0.26&0.05& 0.08& 0.12&0.21& 0.27& 0.62&5.5&45.0&39.9&18.1&26.6&25.8\\
"&4&1&1499&961&998 & 0.78& 0.78& 0.46& 0.04& 0.04& 0.10&0.18& 0.18& 0.45&5.3&78.3&92.2&21.8&29.2&30.6
\enddata
\label{table}
\tablecomments{Statistics listing the distribution of 
 low-mass stars (LM; 0.3--0.5 M$_{\odot}$), intermediate-mass stars (IM; 0.8--1.5 M$_{\odot}$), 
and high-mass stars (HM; 1.5--3 M$_{\odot}$) for the observed population (Obs.) and the two 
model runs (M1 and M2).  "Number" refers to the number of planets with final semimajor axes $<$ 3 AU.
The relative fraction of planets is computed for a$_{p}$$<$0.2 AU, a$_{p}$=0.2--0.5 AU, 
and a$_{p}$=0.5--3 AU for each stellar mass bin.
The $\chi^{2}$ parameter is computed by comparing the relative fraction of simulated planets with the observed fraction 
for each semimajor axis bin for a sample size equal to the number of observed planets in each stellar mass bin.  
The column "$\%$" lists the percentage of gap-opening planets for each stellar mass bin regardless of final semimajor axis.  
Percentages are not calculated for the first set of simulations because in those simulations planets are assumed to form whenever 
there is sufficient mass for them to form.  
\textbf{(a)} -- Shown for reference is the \textbf{absolute} frequency of all 
detected planets from radial velocity surveys \citep{Jj07b} for different stellar mass ranges.}
\end{deluxetable}

\end{document}